# Natural gas shortages during the "coal-to-gas" transition in China have caused a large redistribution of air pollution

Siwen Wang[a], Hang Su[a,*], Chuchu Chen[a], Wei Tao[a], David G. Streets[b], Zifeng Lu[b], Bo Zheng[c], Gregory R. Carmichael[d,e], Jos Lelieveld[a], Ulrich Pöschl[a] and Yafang Cheng[a,*]

[a]Max Planck Institute for Chemistry, Hahn-Meitner-Weg 1, 55128 Mainz, Germany

[b]Energy Systems Division, Argonne National Laboratory, Lemont, IL 60439, USA

[c]Laboratoire des Sciences du Climat et de l'Environnement, LSCE, Bat 714, Pe 2024, Orme de Merisiers, 91191 Gif-sur-Yvette, France

[d]College of Engineering, University of Iowa, Iowa City, IA 52242, USA

[e]Center for Global and Regional Environmental Research, University of Iowa, Iowa City, IA 52242, USA

*To whom correspondence may be addressed. Email: yafang.cheng@mpic.de, or h.su@mpic.de.

## Abstract

The Chinese "coal-to-gas" strategy aims at reducing coal consumption and related air pollution by promoting the use of clean and low carbon fuels in northern China. Here we show that on top of meteorological influences, these measures achieved an average decrease of fine particulate matter ($PM_{2.5}$) concentrations of ~14% during winter 2017 in Beijing and surrounding areas (the "2+26" pilot cities). However, the localized air quality improvement was accompanied by a contemporaneous ~15% upsurge of $PM_{2.5}$ concentrations over large areas in southern China. We find that the pollution transfer that resulted from a shift in emissions was caused by a natural gas shortage in the south due to the "coal-to-gas" transition in the north. The overall shortage of natural gas greatly jeopardized the air quality benefits of the "coal-to-gas" strategy in winter 2017 and reflects structural challenges and potential threats in China's clean energy transition.

**Introduction**

The notorious "airpocalypse" in China stems from a multitude of air pollutants (1–3) that are associated with significant climate and health effects (4–8). Eliminating the severe fine particulate matter ($PM_{2.5}$, particulate matter with a diameter smaller than 2.5 μm) pollution smog has been perceived as a national priority, with the establishment of a coal consumption cap that required the share of coal in the national primary energy mix to drop to below 65% in 2017 by a transition to cleaner natural gas and non-fossil energy sources (9). Coal reduction is, however, the crux of China's air pollution control (10). Fig. 1 shows the coal control roadmap of China from 2010–2030. As of 2017, over half of the world's coal consumption has occurred in China, accounting for ~60% of the country's primary energy consumption (11). After effective coal reductions from the power sector and key energy-intensive industries (Phase I), the further coal control in China has been focussing on the reduction of dispersed coal use in residential and small industrial facilities (Phase II). The dispersed coal (the so-called "Sanmei" in Chinese) refers to raw coal, usually a high-polluting fuel with high-ash residue, burned in non-centralized combustion facilities without end-of-pipe air pollutant treatment. The residential dispersed coal combustion in vast rural areas has been estimated to be a major contributor to high $PM_{2.5}$ exposure and premature mortality in China (7, 12–14).

Since early 2017, a series of clean heating actions, namely "coal-to-gas" and "coal-to-electricity", have been implemented in Beijing and its neighbouring provinces, especially in the "2+26" cities located along the air pollution transport channel of the Beijing-Tianjin-Hebei (BTH) region (Beijing, Tianjin, and other 26 cities in Hebei, Shanxi, Shandong and Henan provinces) (15, 16). The major measures include replacing traditional household coal-fired stoves with wall-mounted natural gas heaters or electric stoves and eliminating the small industrial coal-fired steam boilers and construction materials industrial kilns (brick, ceramic and lime industries) (17). The changes in coal consumption related to individual measures are given in Fig. 1.

To ensure the implementation of the "coal-to-gas" action, the Chinese authorities have increased domestic natural gas production, supplemented by acquisitions from major international exporters. However, we find that the national natural gas consumption has increased by only 2.1 billion cubic metres in volume (bcm) in the last two months of 2017 compared to the same period of 2016 (18), far below the estimated increment required in the "coal-to-gas" action in northern China (3.8–5.1 bcm for two heating months, see *Methods*). A recent policy notice released by the Chinese National Energy Administration (NEA) (19) has also confirmed the nationwide natural gas shortage during the "coal-to-gas" action in winter 2017 (see *SI Appendix*, sections 1 and 3). Since the natural gas supply was made a priority for use in "coal-to-gas" homes in northern China (16), cross-regional transfers from the natural gas quotas for other regions in China was inevitable in that winter, which has consequently led to air pollution transfer to those regions that had to use more polluting energy alternatives due to natural gas shortage.

**Results**

**Improved air quality in the "coal-to-gas" cities.** To assess the overall air quality benefits from the "coal-to-gas" action and other measures regarding dispersed coal reductions in winter 2017, we conducted a synthesis analysis of ground-based measurements of $PM_{2.5}$ and its precursors sulfur dioxide ($SO_2$) and nitrogen dioxide ($NO_2$) (*SI Appendix*, Fig. S1), chemical transport model simulations, and energy-related primary emission-change estimates over China. Ground-based measurements recorded at over 1,500 atmospheric monitoring sites show that the $PM_{2.5}$ concentrations declined dramatically in the heating period 2017 compared to 2016 over almost the entire northern China (*SI Appendix*, Fig. S2). Here, data from the last two months (November–December) of each year are used to represent the heating period in winter. We discard other heating months in the analyses because the "coal-to-gas" action has been temporally halted in early 2018 due to the natural gas shortage (20). The average $PM_{2.5}$ concentrations dropped by 41.8% (53.6 µg m$^{-3}$) over the "2+26" cities and 60.7% (68.9 µg m$^{-3}$) over Beijing. Interestingly, across the North-South Central Heating Supply Line denoted in Fig. 2—a historical boundary from the 1950s that geographically distinguishes the northern territories with central heating supply in winter from southern China—the contemporaneous $PM_{2.5}$ concentrations, however, mostly increased.

To account for the meteorological influence on the interannual variability of air pollution (21–24), GEOS-Chem model simulations were performed for 2016–2018, with anthropogenic emission inputs fixed to the 2016 monthly levels. The model framework and determination of meteorological effects are described in *Methods* with model evaluations on meteorological parameters (*SI Appendix*, Fig. S3) and surface $PM_{2.5}$ (*SI Appendix*, Fig. S4), as well as sensitivity examination for emission intensity and model resolution (*SI Appendix*, Fig. S5). Our modeling results suggest that the meteorological effects contributed an average ~28% decrease of PM$_{2.5}$ (~36 μg m$^{-3}$) in the heating period 2017 relative to the 2016 level over the "2+26" cities in northern China (*SI Appendix*, Fig. S2); and ~40% over Beijing, which is comparable to the ~42% estimation inferred from a recent study on Beijing's PM$_{2.5}$ based on Weather Research and Forecasting (WRF) meteorological inputs (25). The meteorological effects were then subtracted month-by-month from the total measured relative changes of PM$_{2.5}$ concentrations, shown in Fig. 2 *A*, which resulted in a residue of 13.8% (17.7 μg m$^{-3}$) net decrease over the "2+26" cities and thus quantifies the effects from all emission control measures.

We further estimate the emission reductions from the dispersed coal control measures over the "2+26" cities in the heating period 2017 by assembling the sector-specific dispersed coal reduction activities, the converted natural gas and electricity coal consumption, and the average sector- and fuel-specific emission factors (*SI Appendix*, Table S1). It turns out that the control measures led to 43.7 Gg (15.3%), 64.6 Gg (13.8%) and 7.8–9.3 Gg (1.0–1.2%) reductions in primary PM$_{2.5}$, SO$_2$ and NOx emissions (see *Methods* and *SI Appendix*, Fig. S6 for detailed data and step-by-step budgets), respectively, based on total emissions for 2016 from the Multi-resolution Emission Inventory for China (MEIC) inventory (26, shown in *SI Appendix*, Table S2). The emission changes in PM$_{2.5}$ and its precursors are consistent with the concentration changes over these regions, among which the "coal-to-gas" action alone accounts for about 58–64% of the decreased PM$_{2.5}$. These results clearly demonstrate the crucial importance and high efficiency of the clean energy transition to air pollution control.

**Deterioration of air quality in the gas-shortage regions.** However, after accounting for the meteorological effects, the emission-induced changes in PM$_{2.5}$ concentrations showed significant increases in large areas outside the northern "2+26" cities in eastern China during the heating period 2016–2017 (Fig. 2 *A*). The upsurge of ambient PM$_{2.5}$ was especially notable over four inland provinces (including Hubei, Hunan, Anhui, and Jiangxi) adjacent to the North-South Central Heating Supply Line, with an average net increase of 14.5% (9.5 μg m$^{-3}$) during the heating period 2017. A significant north-south shift is observed when comparing the distribution of the adjusted relative changes of ambient PM$_{2.5}$ between these southern 51 cities (red) and the northern "2+26" cities (blue) in Fig. 2 *C*. Note that meteorological effects actually contributed an average decrease of 5.6% in PM$_{2.5}$ over these four southern provinces (*SI Appendix*, Fig. S2). In the heating period 2018, no significant upsurge of $PM_{2.5}$ was found in either the northern or southern regions, when natural gas supply was sufficient (see *SI Appendix*, section 2) (Fig. 2).

We further examine the evolution of the meteorologically adjusted $PM_{2.5}$ and $SO_2$ concentrations in both non-heating (July–October) and heating periods during 2016–2018 for these northern and southern cities in Fig. 3 (see more in *SI Appendix*, Fig. S7). In the non-heating period, these air pollutants show synchronously declining trends over both regions, reflecting the regular progress of emission mitigation throughout China in recent years (for primary $PM_{2.5}$ and $SO_2$, mainly in industrial and residential sectors, see *SI Appendix*, Table S2) (26). However, a distinct south-shifted $PM_{2.5}$ pattern can be distinguished in the heating period 2017 compared to the other years. This south-shift can be more clearly seen in the discrepancies of $PM_{2.5}$ ($\Delta PM_{2.5}$) and $SO_2$ ($\Delta SO_2$) concentrations between the non-heating and heating periods. In contrast to the declining tendencies in $\Delta PM_{2.5}$ and $\Delta SO_2$ in the "2+26" cities in the north, they surged by up to 43.4 µg m$^{-3}$ for $\Delta PM_{2.5}$ and 5.8 µg m$^{-3}$ for $\Delta SO_2$ in the southern 51 cities in the heating period 2017. Again, such a regional air pollution transfer did not occur in winter 2018.

The distinctive regional decoupling of air pollution tendencies in the exceptional heating period of 2017 indicates a dominant role of the cross-regional natural gas transfers on the air pollution shift from the northern "coal-to-gas" regions to other areas of the country. The interrupted natural gas supply developed rapidly in a "light-to-severe" sequence from the use in industrial processes and industrial fuels to natural gas vehicles, public services sectors, urban heating, and residential users (27). Among the gas-shortage areas outside the northern "2+26" cities, the above-mentioned four southern provinces were confronted with the greatest difficulties according to reports from local agencies (18). As an example, the purchase of residential natural gas was limited to 150 m$^3$ per month for each home in Wuhan, the capital of Hubei province—only one-third of the normal gas demands for an urban family in winter (28, *Methods*). The industrial coal consumption (excluding the non-energy coal use) in these four provinces increased by 5.0% in 2017 compared to 2016, in contrast to the 6.0% decreases for the nation's total; the residential part decreased slightly by 1.4%, also lower than the 3.9% national decreases (numbers shown here for annual amounts) (29). This supports the idea that the significant $PM_{2.5}$ upsurge over the four provinces compared to other southern areas (Fig. 2 *A*) may have resulted from a shift in emissions caused by a natural gas shortage in the south due to the "coal-to-gas" action in the north.

For such a short-term unexpected event, the bottom-up emission inventory (e.g., the MEIC) is not able to provide a correct trend analysis due to lack of monthly and local energy consumption data at the moment (see data provided in *SI Appendix*, Table S2), although the progress of sectoral emission controls has been considered in emission factors (26). For this reason, we conducted holistic emission-change estimates for the most likely affected industrial and residential sectors in the four southern provinces to evaluate the influences of the energy transition there. A conservative estimate of 20% natural gas deficits for these regions in the heating period 2017 was made according to natural gas activities reported from local agencies (e.g., ~20% natural gas shortage in Hubei and Hunan provinces; 22–33% in Hefei, the capital of Anhui province; 22% in Jiangxi province; more information in ref. 18), which corresponds to at least a total of 0.8 bcm natural gas in the two heating months (see *Methods*). The alternative energy source to natural gas was expected to be added in the forms of, for feasible examples, coal- and oil-fired boilers in industries and public services sectors (the backup or outdated facilities reserved for emergency use), electric heaters and air conditioners in urban households, and coal- or biofuel-fired stoves in rural areas. Assuming coal as the dominant alternative energy source (other types of fuels, e.g., oil and biofuels, would have more negative impacts on air quality (13)) to compensate for the energy deficits from the 0.8 bcm natural gas shortfall, approximately 1.9 Mt coal would have to be added over the four southern provinces in view of the equivalent heat value (considering a factor of 1.2 for higher thermal efficiency of gas-fired facilities, see *Methods*). The estimated corresponding air pollutant emissions from this additional coal use have large uncertainties, given scarce energy activity and facility information. Nevertheless, it is estimated that the additional coal use would result in net emission increases of 4.5% (11.1 Gg) for primary $PM_{2.5}$ and 7.0% (23.7 Gg) for $SO_2$, if completely burned in residential stoves, or 0.9% (2.3 Gg) for primary $PM_{2.5}$ and 3.8% (12.7 Gg) for $SO_2$, if burned in industrial boilers, based on the sectoral-average abated emission factors adopted from the MEIC inventory for 2016 (*SI Appendix*, Tables S1 and S2). The changes in NOx emissions are negligible. These emission changes are lower than the 14.5% of ambient $PM_{2.5}$ upsurge over these southern four provinces. However, it should be noted that our estimate on the additional coal use for the four southern provinces is on the conservative side (e.g., the natural-gas-shortage amounts, neglect of oil and biofuel as alternative energy sources, and thermal efficiency discrepancy between gas-fired and coal-fired facilities). Also, it can be expected that the contingency industrial boilers and residential stoves were mostly in low maintenance and non-optimal operational status or phased out by recent environmental regulation, and their air pollutant emission factors could be as much as 2–8 times larger (30, 31) (or higher without end-of-pipe controls) than those used in the MEIC inventory for 2016 which represent the already-improved emission control levels under strict environmental management in recent years. In the industrial processes for alumina and non-ferrous metals production, the emission factors for $PM_{2.5}$ from coal combustion can even be 50 times higher (31). Thus, there is a high likelihood that the additional air pollutant emissions from the alternative fuels due to natural gas shortage ("gas-to-others") played a major role in deteriorating the air quality over these four provinces in southern China.

**Discussion and Policy Implications**

Fig. 4 integrates the changes in air quality and energy supply indices in the heating period 2016–2017 for the two specific regions in northern and southern China. The regional air pollution redistribution, triggered by the "coal-to-gas" action in northern China in the context of the severe natural gas shortage, may greatly jeopardize the overall air quality benefits (almost completely offset) to be expected from the transition toward cleaner energy and environmental justice. However, given sufficient natural gas supply as in winter 2018, such regional air pollution transfer can be avoided and a nationwide decline of ambient $PM_{2.5}$ concentrations can be achieved, as shown in Fig. 2 *B*. Our findings highlight the necessity of a coordinated framework for the energy-environment nexus, where the potential co-benefits for climate and human health should be fully considered from local to regional scales (32).

China is implementing its energy transition under a tight schedule. In 2030, the shares of natural gas and non-fossil energy are estimated to increase to 15% and 20% of the primary energy mix, respectively (33) (Fig. 1). By that time, the importance of natural gas as part of the regional air quality controls will be further highlighted. However, there is increasing recognition that the demand-supply imbalances of natural gas will significantly widen over the Asia-Pacific region in the next decade (34), and a repetition of the severe natural gas shortage of winter 2017 and the consequent redistribution of air pollution will still be a high risk in China. Concerning China's upsurge of natural gas demands and national energy security (*SI Appendix*, section 4), coordinated efforts that strive for the introduction of new and improved clean technologies and the effective utilization of coal, gas and alternative energy sources are needed to sustain the national and regional energy demands while achieving air quality goals. The risks and uncertainties in China's clean energy transition are also worthy of consideration by policymakers in other coal-dominated countries. An efficient and well-coordinated energy transition in China will benefit not only regional air quality and human health but also the sustainable development of the world.

**Materials and Methods**

**Air pollution measurements.** The hourly ground-based measurement data for ambient $PM_{2.5}$ and its precursors $SO_2$ and $NO_2$ concentrations are recorded at more than 1,500 atmospheric monitoring sites managed by the Ministry of Ecology and Environment (MEE) of China since 2013. These urban stations geographically cover all prefecture-level districts of mainland China following the requirements of the national technical regulation for ambient air quality monitoring (35). Each monitoring site reports the average concentrations of air pollutants over the surrounding 0.5–4.0 km (in radius) areas. The minimum number of monitoring sites for a single city depends on its urban population and size, and for this study we applied data from 158 and 255 sites located within the studied "2+26" cities in northern China and the four inland provinces (including Hubei, Hunan, Anhui, and Jiangxi) in southern China (Fig. 2), respectively. The 24-hour average concentrations of $PM_{2.5}$, $SO_2$, and $NO_2$ were used for the air pollution analyses during 2016–2018.

**Model configuration and evaluation.** The GEOS-Chem model (version v11-01) simulations were conducted for 2016–2018 to determine the meteorological effects on the surface air pollution changes over eastern China. GEOS-Chem is a global chemical transport model for air composition studies that simulates multiple physical and chemical processes, driven by assimilated meteorological data from the Goddard Earth Observing System (GEOS) of the National Aeronautics and Space Administration (NASA)/Global Modeling and Assimilation Office (GMAO), with 72 hybrid sigma-pressure levels in the vertical (14 layers in the lowest 2 km) extending up to 0.01 hPa. The simulation was performed at 2.5°longitude × 2°latitude horizontal resolution with 47 vertical layers (in reduced mode), compiled with the "SOA" chemical mechanism that includes the secondary organic aerosol (SOA) formation. The MERRA-2 reanalysis meteorological data (updated every 1–3 hours) was used in this study, which covers a continuous time period from the year 1980 onwards, and has been widely evaluated on multiple parameters using satellite measurements (36–39). At least one-year spin-up simulation was conducted to minimize the effects from initial concentration fields.

The global anthropogenic emissions of atmospheric species were provided by the Emission Database for Global Atmospheric Research (EDGAR v4.2) inventory for 2012 (40, 41), and was replaced over China by the MEIC inventory (v1.3) for 2016 (26, 42–44). The MEIC database includes monthly air pollutant emissions for $SO_2$, NOx, carbon monoxide (CO), ammonia ($NH_3$), volatile organic compounds (VOCs), primary $PM_{2.5}$ and coarse PM, black carbon (BC) and organic carbon (OC) at 0.25° × 0.25° horizontal resolution (http://www.meicmodel.org/). The international ship emissions for CO and NOx were provided by the International Comprehensive Ocean-Atmosphere Data Set (ICOADS) inventory (45), and for $SO_2$ from the Arctic Research of the Composition of the Troposphere from Aircraft and Satellites (ARCTAS) inventory based on the work by Eyring *et al*. (46, 47). The aircraft emissions of NOx were from the Aviation Emissions Inventory Code (AEIC v2.1) inventory (48–50). Global Fire Emissions Database (GFED v4) was used for biomass burning emissions (51). Biogenic VOCs and NO emissions were calculated by the Model of Emissions of Gases and Aerosols from Nature (MEGAN v2.1) (52). Other natural source emissions include the soil and lightning emissions for NOx (53, 54), and the volcanic $SO_2$ emissions from AeroCom point source data (55). The daily 24-hour average concentrations of $PM_{2.5}$ and precursor gases were derived for the surface layer of the model, which typically has a height of about 120 m in winter over eastern China.

The meteorological parameters used in the model, including air temperature (at 2 m height), relative humidity (RH) and wind speed, were evaluated using the NCEP ADP Global Surface Observational Weather Data (https://rda.ucar.edu/datasets/ds461.0/) (Supplementary Fig. 3). Statistical parameters were calculated for correlation coefficient (R), mean bias (MB), root mean square error (RMSE), normalized mean bias (NMB), and normalized mean error (NME). Results suggested that the MERRA-2 data reproduced well the meteorological variations in the second half of 2016–2018 over eastern China. No significant regional discrepancy was found. The simulated surface $PM_{2.5}$ concentrations were compared with the MEE measurements for the second half of 2016 over eastern China (*SI Appendix*, Fig. S4), and showed a high correlation (R = 0.83) with no significant biases (NMB = 0.5%).

To examine the role of model resolution on the estimates of meteorological effects, we also conducted the one-way nested GEOS-Chem simulations over China and Southeast Asia (60°E–150°E, 11°S–55°N) for 2016–2018 (details see refs. 56, 57). The nested GEOS-Chem model was performed at 0.625°×0.5° horizontal resolution with 47 vertical layers, driven by MERRA-2 meteorological data. Similar emission inputs were used following that of the global GEOS-Chem simulations.

**Meteorological effects and adjustments.** The observed relative changes of air pollutant concentrations between two time periods are linearly decomposed into two pieces of sources: meteorology-induced (MI) and emission-induced (EI) relative changes (58). The former is determined by the meteorological factors and chemical mechanism in the GEOS-Chem model with a fixed setting of emission inputs. Here, the year 2016 is selected as a basis for calculation. The meteorology-induced relative change in the observed concentration ($\Omega_{obs}$) of year $i$ is considered to be a first approximation of the modeled ($\Omega_{mod}$) relative change:

$$\text{MI} = (\Omega_{adj,i} - \Omega_{obs,i}) / \Omega_{obs,2016} = (1 - \Omega_{mod,i} / \Omega_{mod,2016}) \tag{1}$$

where the $\Omega_{adj,i}$ presents the adjusted observed concentrations of year $i$ with the compensatory of meteorology-induced piece of concentration changes ($\Omega_{obs,2016} \times \text{MI}$). This approach has the advantage of yielding local meteorological effects that account for the nonlinear impacts over a variety of terrain. It then derives the adjusted observed concentrations of year $i$:

$$\Omega_{adj,i} = \Omega_{obs,i} + \Omega_{obs,2016} \times (1 - \Omega_{mod,i} / \Omega_{mod,2016}) \tag{2}$$

The difference between the meteorology-induced relative change and the total relative change in the observed concentration of air pollutants in year $i$ is attributed to the emission-induced part:

$$\text{EI} = \Omega_{mod,i} / \Omega_{mod,2016} - \Omega_{obs,i} / \Omega_{obs,2016} \tag{3}$$

Similar evaluation of meteorological effects has been widely applied (22, 23, 25, 58–60). The potential biases of this approach are mainly associated with the model performance on reproducing the real meteorological fields and the air pollutant concentrations, although the model used here is considered a state-of-the-art tool and data source for the external adjustments on (or comparators to) ground-based measurements (61–64). Sensitivity examination showed that using different anthropogenic emission inputs (e.g., the MEIC inventory for 2010, with 45.7% and 107.5% higher primary $PM_{2.5}$ and $SO_2$ emissions than 2016) merely produced a small influence (relative error less than 10%) on the meteorological effect estimates for both studied regions. The different scales of the model and ground-based measurements can also introduce systematic biases. However, similar meteorological effect estimates were generated—erasing 31.6% and 8.1% $PM_{2.5}$ concentrations over the northern

"2+26" cities and four southern inland provinces, respectively—using the regional-scale nested GEOS-Chem simulations (*SI Appendix*, Fig. S5). These indicated that the above-mentioned systematic and random biases can be well reduced via regional averaging and comparison between relative changes. The tendencies of $PM_{2.5}$ changes are robust over the studied two regions in this study.

**Emission-change estimates for major air pollutants.** The emission changes of $PM_{2.5}$ and its precursors $SO_2$ and NOx were estimated for the "2+26" cities in northern China and the four inland provinces in southern China. The changes in the ambient concentrations of $PM_{2.5}$ and satellite-observed AOD (aerosol optical depth) were found tightly correlated with the changes in primary $PM_{2.5}$ and $SO_2$ emissions (26, 65). Thus, the estimated changes in emissions of primary $PM_{2.5}$ and $SO_2$ should to a large extent reflect the changes in ambient $PM_{2.5}$ concentrations. Although the MEIC inventory provides monthly air pollutant emissions up to 2017, the downscaled activity data on the national level for 2017 and the annual-based provincial energy consumption data can introduce large biases to the two-month emission-change analyses over the specific regions in this study (26). Instead, we used the sector-specific dispersed coal reduction data provided by the China Dispersed Coal Governance Report (17) and the converted natural gas consumption for the "2+26" cities in northern China. Lacking monthly energy data at a provincial level (similar bottleneck for the MEIC inventory), the changes in energy activities for the four southern provinces were based on reports from local agencies on the natural gas shortage (18) and varied by possible energy transition ("gas-to-others") in different sectors. The sector- and fuel-specific emission factors for the two regions were adopted from the MEIC inventory for 2016 (*SI Appendix*, Table S1). The total MEIC emissions for the heating period 2016 were used as the basis for the regional relative emission change analyses (*SI Appendix*, Table S2).

According to the China Dispersed Coal Governance Report (17), the clean heating measures have prevented a total of 65.4 Mt dispersed coal use over the BTH and surrounding areas in 2017, including 18.0 Mt from the residential clean heating in about 6 million homes (4.75 million homes in the "2+26" cities, about 70% transited to natural gas and 30% to electricity), and 15.0 and 32.4 Mt due to the phaseout of small industrial boilers and construction materials industrial kilns.

There are typically four heating months during winter in most areas in northern China, except for the Northeast region (six months). The residential coal consumption for heating is about 3.0 tonnes per year (i.e. 0.75 tonnes per heating month) in each home. For an urban family, a single commercial wall-mounted natural gas heater usually consumes 12.0–15.0 m$^3$ natural gas per heating day (66), with the upper bound for cold indoor temperature or cold walls. For rural families, the natural gas use was reported over 20.0 m$^3$ per heating day (67). We adopted the 15.0–20.0 m$^3$ daily natural gas use per "coal-to-gas" home in this study, which yields 3.8–5.1 bcm natural gas consumption in two heating months for the 4.2 million "coal-to-gas" homes in northern China.

In the "2+26" cities, the involved households are 4.75 million homes, corresponding to 7.1 Mt dispersed coal reductions (5.0 Mt by "coal-to-gas" and 2.1 Mt by "coal-to-electricity") and 3.0–4.0 bcm natural gas increment during the heating period 2017. The phaseout of small industrial boilers and kilns accounts for 44.0% and 36.5% of the total numbers in the BTH and surrounding areas, respectively (17), contributing 3.1 Mt dispersed coal reductions in the two-month period. Thus, the reduced air pollution emissions are 41.5 Gg for primary PM$_{2.5}$, 50.1 Gg for SO$_2$, and 6.5 Gg for NOx (the mass calculated as NO$_2$) from the residential dispersed coal reductions, and 2.9 Gg for PM$_{2.5}$, 15.9 Gg for SO$_2$, and 10.1 Gg for NOx from the phaseout of small industrial facilities. The additional 3.0–4.0 bcm natural gas consumption produces 4.4–5.9 Gg NOx and negligible PM$_{2.5}$ and SO$_2$. Considering the equivalent 2.1 Mt residential coals avoided by "coal-to-electricity" as combusted in local coal-fired power plants, the corresponding emissions are 0.7 Gg for PM$_{2.5}$, 1.4 Gg for SO$_2$, and 2.9 Gg for NOx. In summary, the net emission reductions due to these measures are estimated at 43.7 Gg (15.3%) for PM$_{2.5}$ and 64.6 Gg (13.8%) for SO$_2$ in the "2+26" cities in the heating period 2017 compared to the same period in 2016, while the reductions in NOx are relatively small (7.8–9.3 Gg, 1.0–1.2%) due to the higher combustion temperature of natural gas than coal. The avoided dispersed coal consumption through "coal-to-electricity" may also be compensated for through the electricity transmission from other provinces (68, 69), and excluding the corresponding emission increment will not cause significant biases in this study. The step-by-step emission-change budgets for these two areas are provided in *SI Appendix*, Fig. S6.

In the four provinces in southern China, we made a conservative estimate of 20% natural gas deficits in the heating period 2017 according to reports from local agencies (see detailed materials in ref. 18). The total annual natural gas consumption (including the liquefied natural gas, LNG) for these provinces was 14.3 bcm in 2017 (29). The fraction of natural gas consumption for the last two months in annual totals was calculated as the national average (23.7%) for 2016 (18). Hence, at least a total of 0.8 bcm natural gas deficit was estimated in the heating period 2017, corresponding to 1.9 Mt coal in view of the equivalent heat value (1 m$^3$ natural gas equals to 1.862 kg raw coal (29), with a factor of 1.2 for higher thermal efficiency of gas-fired facilities (70)). The air pollution emissions from the additional coal are estimated at 11.1 Gg (4.5%) for PM$_{2.5}$ and 23.7 Gg (7.0%) for SO$_2$, if burned in residential stoves, or 2.3 Gg (0.9%) for PM$_{2.5}$ and 12.7 Gg (3.8%) for SO$_2$, if burned in industrial boilers, and negligible changes for NOx.

**Data availability**

The ground-based air quality data from the MEE monitoring stations used to support the major findings of this study are accessible from https://beijingair.sinaapp.com/.

**Acknowledgments**

This work is supported by the Max Planck Society (MPG). Y.C. acknowledges the Minerva Program of MPG.

**Author Contributions**

H.S. and Y.C. conceived the study. S.W. performed the research, including data analyses, model simulation and emission estimates. C.C. supported the data visualization. W.T. supported the meteorology validation. B.Z. provided the MEIC inventory and regional emission factors. D.G.S. and Z.L. checked emission estimates. D.G.S., Z.L., G.R.C., J.L. and U.P. commented on the results and manuscript. S.W., Y.C. and H.S. wrote the paper with contributions from all co-authors.

28. Sina News, Wuhan limits the purchase of natural gas to 150 cubic metres per month each family (2017). http://finance.sina.com.cn/china/2017-12-21/doc-ifypxrpp3133492.shtml. Accessed 10 January 2020.

29. National Bureau of Statistics, *China Energy Statistical Yearbook 2018* (China Statistics Press, Beijing, 2019).

30. M. Li *et al.*, Anthropogenic emission inventories in China: a review. *Natl. Sci. Rev.* **4**, 834–866 (2017).

31. Y. Lei, Q. Zhang, K. B. He, D. G. Streets, Primary anthropogenic aerosol emission trends for China, 1990–2005. *Atmos. Chem. Phys.* **11**, 931–954 (2011).

32. J. Lelieveld, P. J. Crutzen, Indirect chemical effects of methane on climate warming. *Nature* **355**, 339–342 (1992).

33. National Development and Reform Commission of China, Energy production and consumption transition strategy (2016–2030) (2016). http://www.ndrc.gov.cn/zcfb/zcfbtz/201704/t20170425_845284.html. Accessed 10 January 2020.

34. *Energy Outlook 2035* (British Petroleum: London, 2014).

35. Ministry of Environmental Protection of China, *Technical regulation for selection of ambient air quality monitoring stations (on trial)* (China Environmental Science Press, Beijing, 2013).

36. R. H. Reichle *et al.*, Assessment of MERRA-2 land surface hydrology estimates. *J. Clim* **30**, 2937–2960 (2017).

37. K. Wargan *et al.*, Evaluation of the ozone fields in NASA's MERRA-2 reanalysis. *J. Clim* **30**, 2961–2988 (2017).

38. Y.-K. Lim, R. M. Kovach, S. Pawson, G. Vernieres, The 2015/16 El Niño event in context of the MERRA-2 reanalysis: A comparison of the tropical pacific with 1982/83 and 1997/98. *J. Clim* **30**, 4819–4842 (2017).

39. V. Buchard *et al.*, The MERRA-2 aerosol reanalysis, 1980 onward. Part II: Evaluation and case studies. *J. Clim* **30**, 6851–6872 (2017).

40. J. G. J. Olivier, A. F. Bouwman, C. W. M. van der Maas, J. J. M. Berdowski, Emission database for global atmospheric research (Edgar). *Environ. Monit. Assess.* **31**, 93–106 (1994).
16

**Figures**

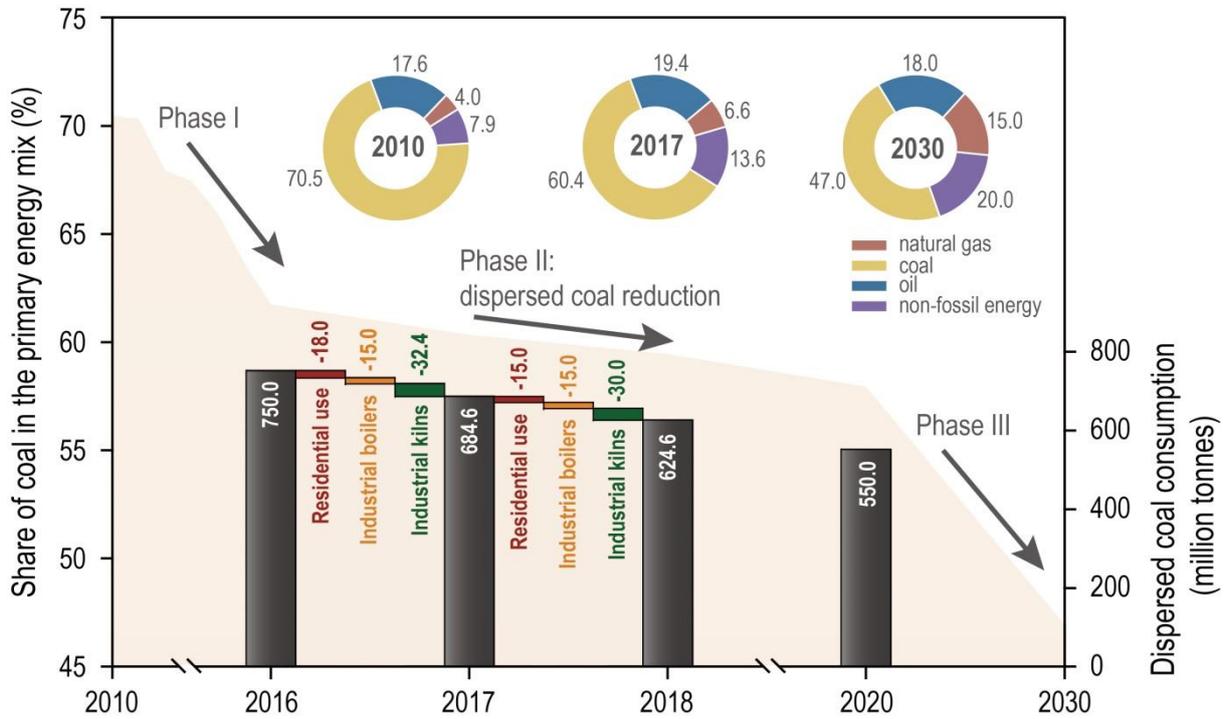

**Figure 1.** The roadmap of coal control in China from 2010–2030. The light orange shaded area shows the share of coal in the primary energy mix for 2010–2030 (11) with three-phased coal control (Phase I: in the power section and key energy-intensive industries; Phase II: toward dispersed coal reductions; and Phase III: pertaining to clean energy development). The black bars indicate the dispersed coal consumption for 2016–2018 and 2020 with source decomposition for sectional changes (17). The doughnut charts show the primary energy structure of China in 2010, 2017 and 2030, respectively (11, 33). Energy data for Hong Kong, Macau and Taiwan are not included.

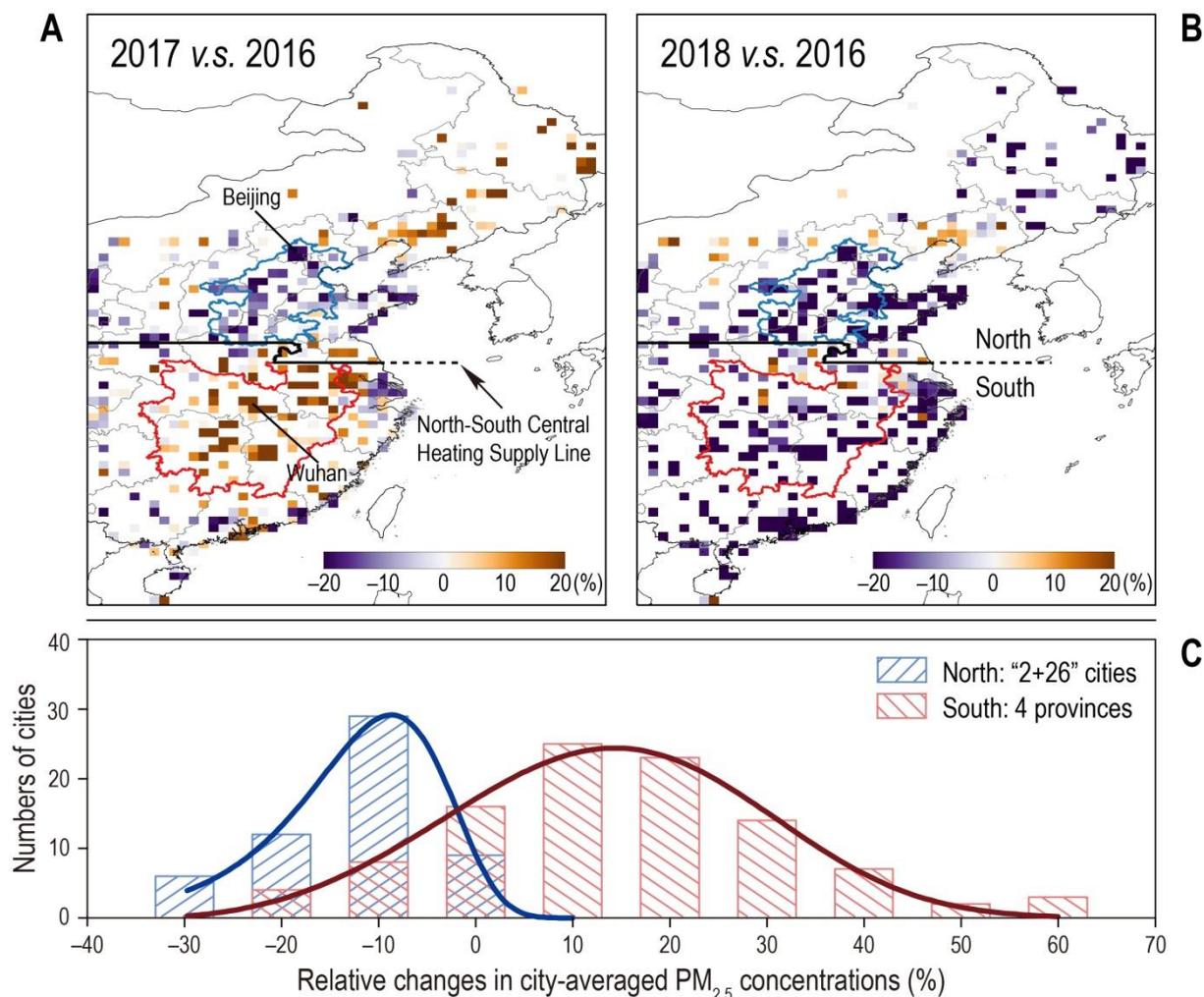

**Figure 2.** Significant north-south regional discrepancies in ambient PM$_{2.5}$ changes during the heating period 2017. (*A* and *B*) The emission-induced relative changes of ground-based PM$_{2.5}$ concentrations for the heating period (November–December) of 2017 (*A*) and 2018 (*B*) contemporaneously compared to 2016 with meteorological effect adjustment. (*C*) The distribution of adjusted PM$_{2.5}$ changes for the studied northern (28 cities, marked with a blue solid line in *A* and *B*) and southern (51 cities, marked with a red solid line in *A* and *B*) regions. Black solid lines in *A* and *B* represent the North-South Central Heating Supply Line. Vertical bars in *C* show the numbers of cities with decreased (negative) or increased (positive) PM$_{2.5}$ concentrations in November and December 2017 for both regions; colored curves represent fitting curves with a Weibull function.

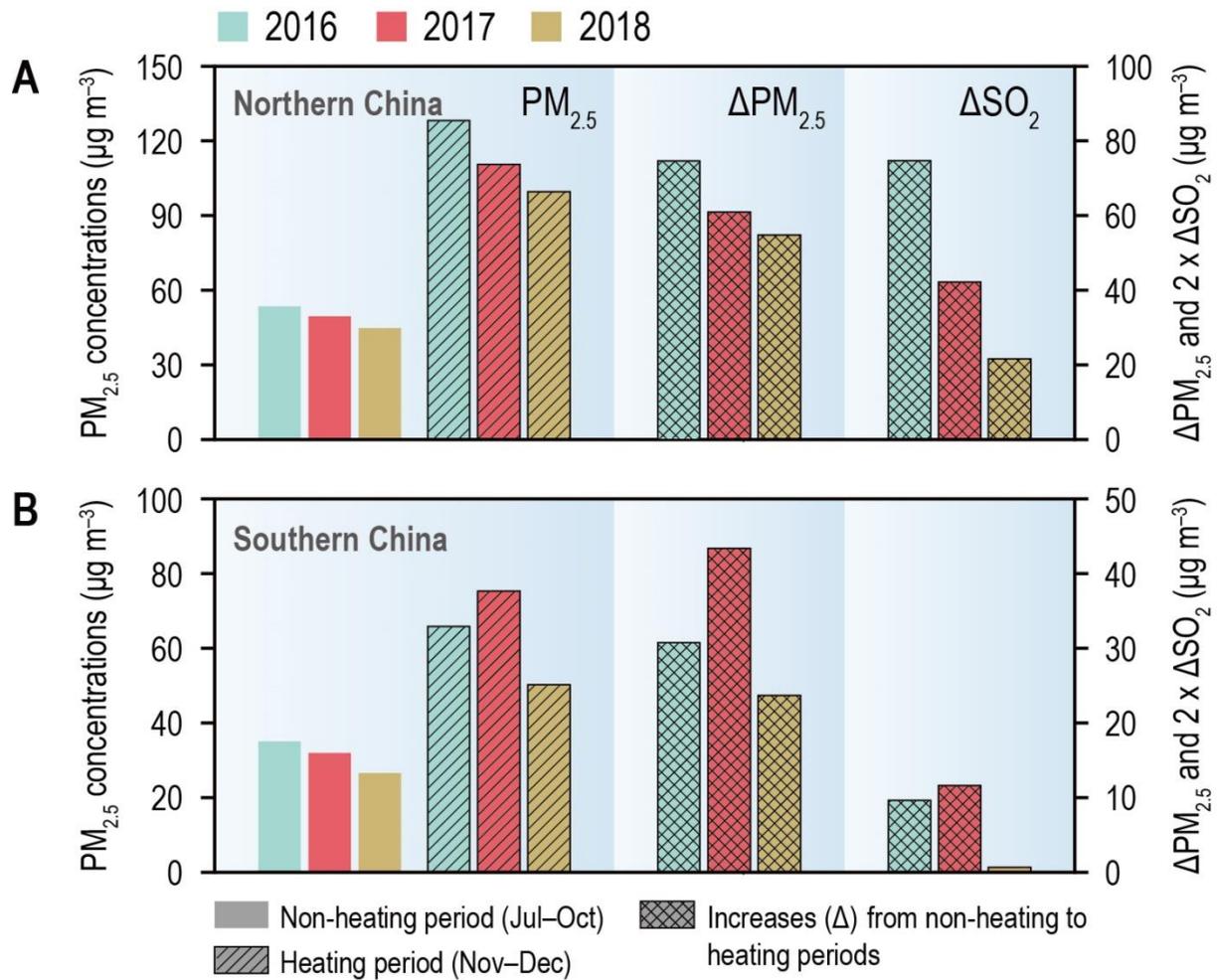

**Figure 3.** South-shifted air pollution pattern in the heating period 2017. The left panel shows the average PM$_{2.5}$ concentrations over the studied regions in northern (*A*) and southern (*B*) China in non-heating and heating periods during 2016–2018. The right two panels show the increases in PM$_{2.5}$ (ΔPM$_{2.5}$) and SO$_2$ (ΔSO$_2$) from non-heating to heating periods. The interannual changes of meteorological effects on air pollutants were adjusted based on 2016.

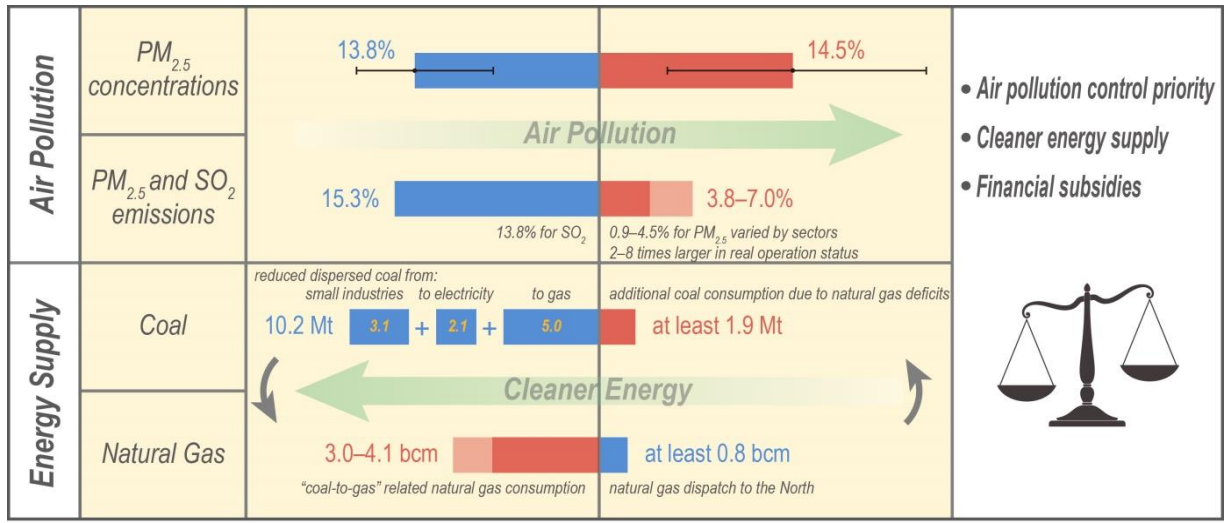

**Figure 4.** The policy determinants of the regional redistribution of air pollution in China. Horizontal bars show the changes in air pollution and energy supply indices during the heating period 2016–2017 over the studied northern "2+26" cities and southern four provinces. The blue and red colors denote negative (decreases in 2017) and positive (increases in 2017) changes, respectively. Error bars for $PM_{2.5}$ concentrations denote the first and the third quartiles. Units of coal, million tonnes (Mt); natural gas, billion cubic metres in volume (bcm).

**Supplementary Information Text**

**Natural gas shortage in winter 2017.** The severe natural gas shortage occurred in winter 2017 attributed not only to the demands for natural gas raised by the "coal-to-gas" action in about 4.2 million homes in northern China, but was also due to several objective reasons, such as (i) the pipeline natural gas import from Turkmenistan was substantially lower than the contract supply amounts (1), (ii) the designed operation of a Tianjin's liquefied natural gas (LNG) receiving terminal owned by Sinopec was not completed as planned (2), and (iii) the weak natural gas infrastructure could not meet the vast cross-regional gas assignment in the peak period. A recent policy notice released by the Chinese National Energy Administration (NEA) has confirmed the nationwide natural gas shortage during the "coal-to-gas" actions in winter 2017 (see more in the later section "Recent policy shift on 'coal-to-gas' in China").

Required by the "coal-to-gas" action, traditional household coal-fired stoves for cooking and heating in most homes had been removed and replaced by natural gas heaters and electric stoves before the heating period 2017. Meanwhile, a coal ban had been imposed in Beijing and surrounding areas in conjunction with the implementation of the clean heating actions (3). Therefore, when the severe natural gas shortage spread nationwide, the natural gas supply had to be made a priority to the northern "coal-to-gas" homes to meet their primary heating demands.

The cross-regional natural gas transfer from the natural gas quotas for other regions (e.g., the southern and northeastern provinces) was inevitable. Natural gas use in most sectors in southern China was greatly affected. For example, the natural gas supply for industrial production activities and the industrial, commercial and public services building heating were disrupted in Wuhan (4). More than 1,700 compressed natural gas (CNG) taxis in Guiyang were out of natural gas supply (5). The household natural gas supply was also affected. For example, the purchase of natural gas was limited to 150 $m^3$ per month for each family in Wuhan (6), only one-third of the normal gas demands for an urban family in winter (~450 $m^3$ per heating month, see Materials and Methods). Though it is difficult to get accurate dispatched amounts of natural gas across provinces due to lack of energy data, the four southern inland provinces seemed to have experienced the greatest gas shortage according to local reports (7).

**"Coal-to-gas" action in winter 2018.** The clean heating actions continued in 2018 in northern China, with another annual 15.0 Mt residential dispersed coal reductions as planned (8). In contrast to the severe natural gas shortage in winter 2017, the natural gas supply was sufficient and stable in 2018 (9). The national total natural gas consumption reached 280.3 bcm in 2018, increased by 18.1% from the 2017 level (7). In the last two months of 2018, the national total natural gas consumption increased by 4.7 bcm compared to that in 2017 (7), closed to the maximum gas demands (5.1 bcm) from the residential "coal-to-gas" in the "2+26" cities contemporaneously in 2017. No significant natural gas shortage has been reported within the country in winter 2018.

The ground-based measurements indicated more significant nationwide declining $PM_{2.5}$ concentrations during the heating period 2018 (19.5%) than 2017 (2.1%) when compared to the 2016 level (*SI Appendix*, Fig. S2 *A,B*). The average $PM_{2.5}$ concentrations over the "2+26" cities in northern China and the four provinces in southern China decreased by 22.3% and 23.6%, respectively, with the meteorological effects adjustments. The $PM_{2.5}$ decreases in winter 2018 over the "2+26" cities have significantly surpassed that in winter 2017 (at 13.8%), which is in accord with the further dispersed coal reductions in 2018 (Fig. 1). The declining $PM_{2.5}$ over the four provinces in southern China together with most other parts of China were also notable in winter 2018, possibly due to the nationwide emission reductions (e.g., 8.1–14.8% reductions in cement production in the studied four southern provinces (10)), due to the slowdown of construction activities in China and the upgraded Sino-US trade friction since 2018.

**Recent policy shifts on "coal-to-gas" in China.** On July 3, 2019, the NEA released a policy notice (11) to solicit comments on the potential issues related to the implementation of clean heating actions during the past two winters in northern China. It confirmed the nationwide natural gas shortage in winter 2017 and the related public awareness on natural gas supply for the "coal-to-gas" actions.

Apart from that, it especially focused on the promotion of clean coal central heating in urban areas and biofuel heating in rural areas—a big shift from the policy priority of "coal-to-gas" in northern China (3). This policy-orientation was widely interpreted as a result of the deteriorating external dependence on natural gas and the financial pressure on energy subsidies for the "coal-to-gas" homes (12, 13). Natural gas is usually regarded as an effective bridge from coal or oil to clean renewable energy sources. The "coal-to-gas" and similar coal-reduced clean energy transition should not be entirely suspended in the North China Plain (NCP) regions, considering their importance to the $PM_{2.5}$ smog mitigation.

**Energy security in China.** China is pursuing its energy transition within a tight schedule compared to the precedents in developed countries. In 2017, the natural gas share was only 6.6% of China's primary energy mix (Fig.1), much lower than the world average (23.4%) (14). To meet the 15% share of natural gas before 2030 set in the national long-term energy strategy (15), China is promoting infrastructure such as a natural gas pipeline network linking to the domestic gas fields, the major international gas providers (East and Central Asia, Myanmar and Russia), and the LNG receiving terminals along the coast (16–18). China has surpassed Japan as the world's largest natural gas importer in 2018, after becoming the world's largest crude oil importer in 2017 (19). The maximum natural gas import is estimated at ~200.0 bcm per year in the period 2025–2030 (120.0 bcm through pipeline import), and by that time China's external dependence on natural gas will increase to about 50.0% (45.3% in 2018) (18). China's natural gas trade will contribute further risks to the international energy market and domestic energy security, as well as air pollution control progress. These kinds of risks are also worth being considered by policy-makers from other coal-dominated countries such as Poland (47.7% coal share in energy mix), Kazakhstan (53.7%), South Africa (68.2%) and India (56.3%) (14). China must find solutions to sustain its energy diversity and strive for breakthroughs in the development of clean and effective utilization of coal, gas, as well as renewable energy technologies, with upgrades in the energy-intensive industries.

Regardless of the natural gas or other non-fossil energy supply, clean energy should be affordable. This is not only meaningful to the market competitiveness of the clean energy, but also important to the sustainable performance of energy transition in low-income rural areas. Although being the world's second-largest economy, China remains a developing country with a substantial population fraction living below the poverty threshold (20). In many rural areas, cheap fuels are the necessities in winter. It remains to be seen if the "coal-to-gas" action can be conducted as a long-term policy measure in China if without the governmental subsidies on the facilities and energy (21).

In contrast to traditional coal utilization, China has the world cleanest coal-fired power plant equipped with the so-called "ultra-low emission" technologies (22), and the world's largest clean-coal power generation system (23). The clean and effective utilization of coal seems technologically promising and cost-effective. China's air pollution roots in its coal-dominated power production structure and should be addressed in less coal-centric (energy diversity) and more energy-environment integrated policies.

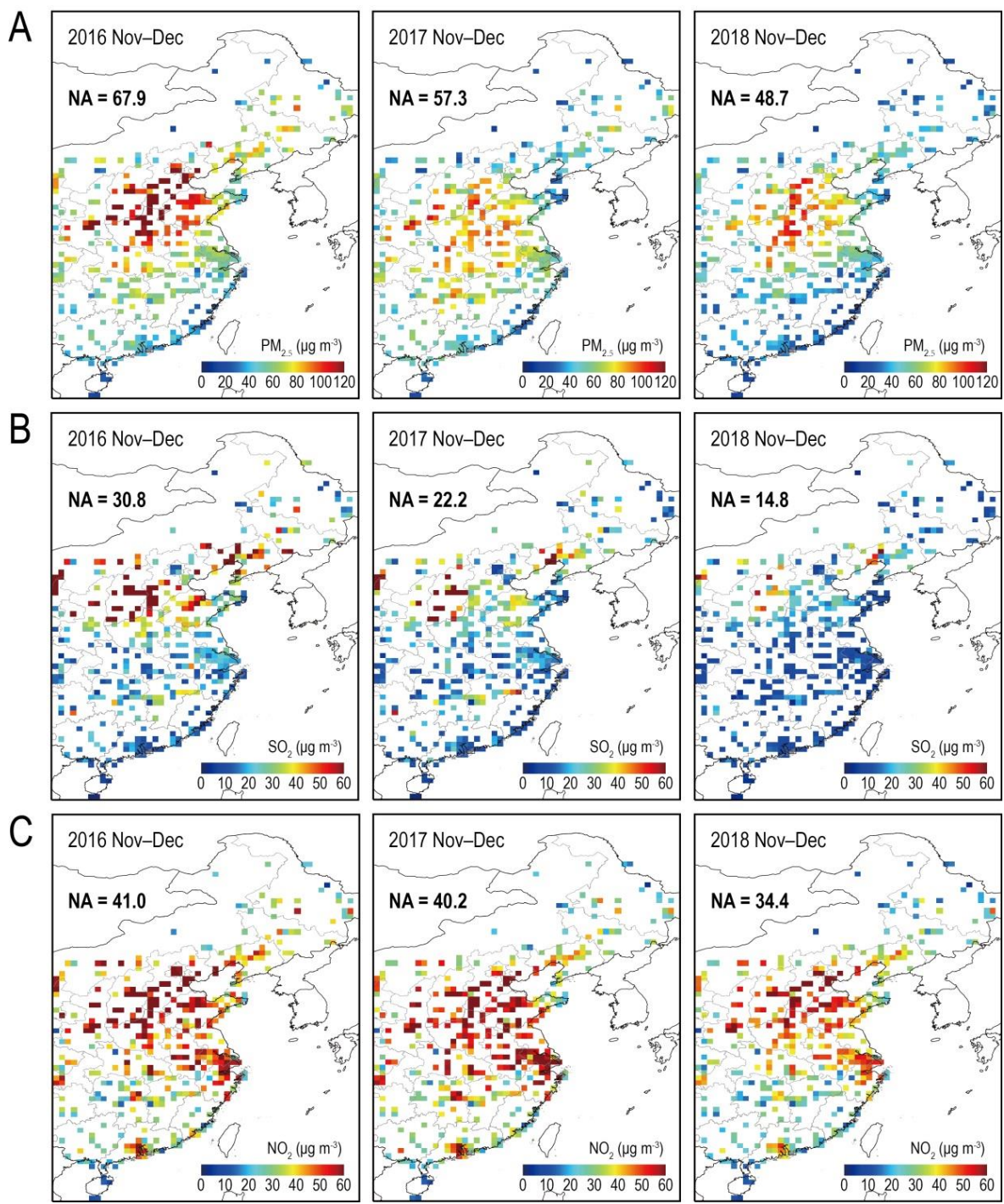

**Fig. S1.** Ground-based measurements of air pollutants in eastern China for the heating period in 2016–2018. **(A–C)** Ambient concentrations of $PM_{2.5}$ (**A**), $SO_2$ (**B**) and $NO_2$ (**C**) for November–December of each year, with the national averages (NA). Data are averaged into $0.5° \times 0.5°$ grids for visual convenience.

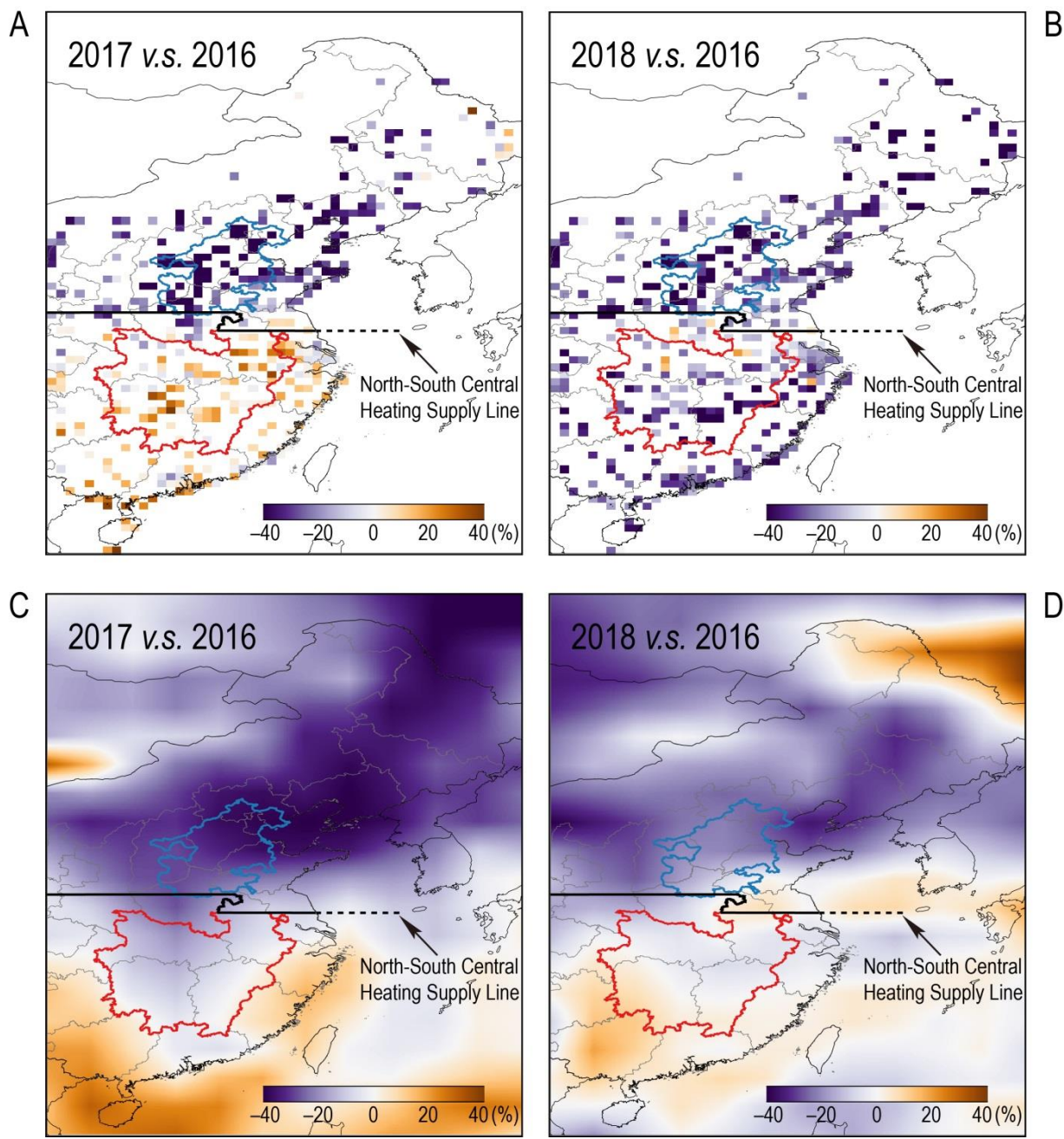

**Fig. S2.** The observed and modeled changes in PM$_{2.5}$ concentrations. Maps show the relative changes of ground-based PM$_{2.5}$ concentrations for the heating periods of 2017 (**A**) and 2018 (**B**), and of GEOS-Chem-simulated PM$_{2.5}$ concentrations (2.5 ° × 2 °, surface layer) for the heating period of 2017 (**C**) and 2018 (**D**), on the basis of the same period of 2016. Symbols, see **Fig. 2** for an explanation.

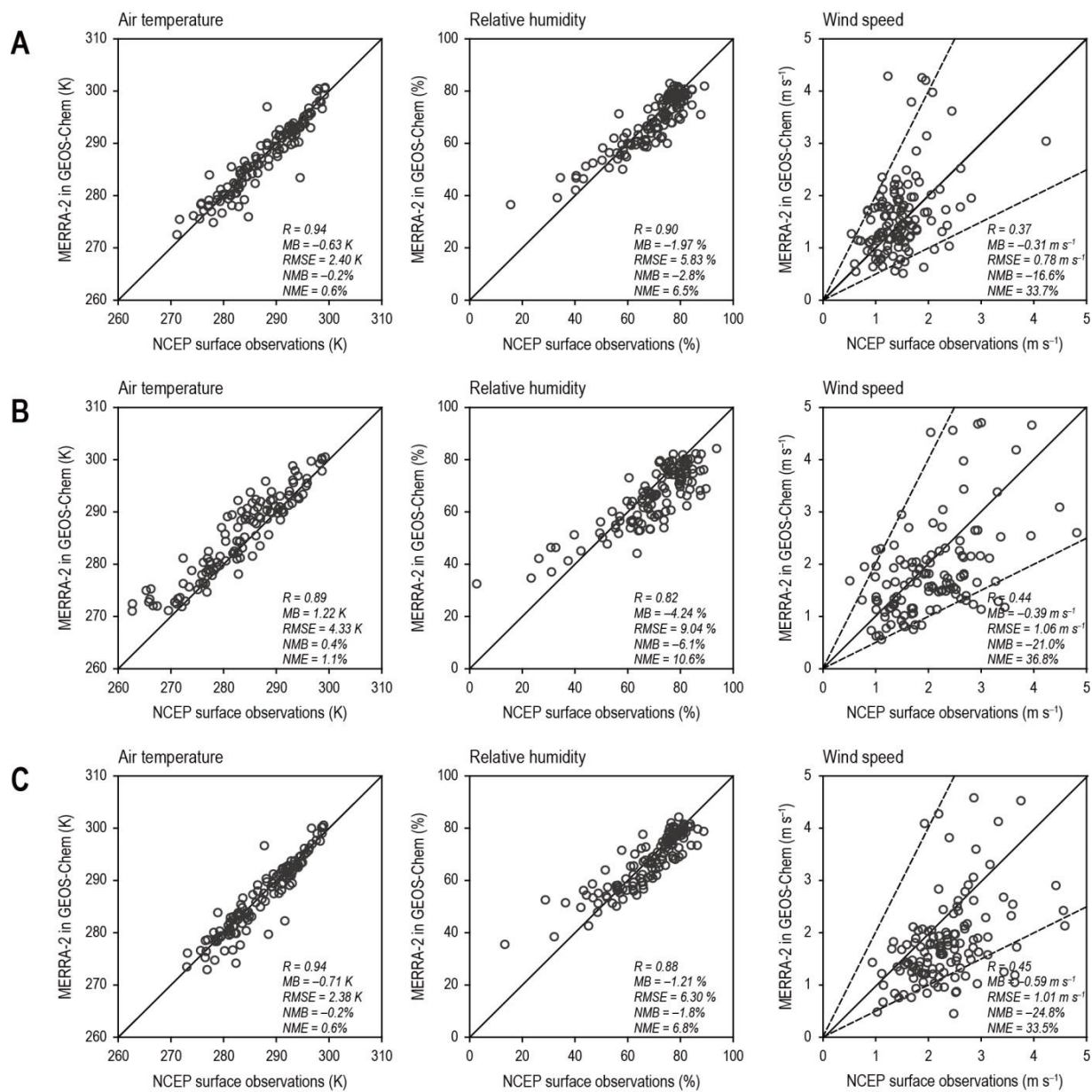

**Fig. S3.** The MERRA-2 meteorological parameter evaluation. Parameters of air temperature (at 2 m height), relative humidity (RH) and wind speed from MERRA-2 were compared to the NCEP ADP Global Surface Observational Weather Data for GEOS-Chem grids over eastern China for the second half of 2016 (**A**), 2017 (**B**) and 2018 (**C**). RH for NCEP data was calculated based on the air temperature and dew point temperature. Statistical parameters for all samples: correlation coefficient (R), mean bias (MB), root mean square error (RMSE), normalized mean bias (NMB), and normalized mean error (NME). The solid lines represent the 1:1 line; dashed lines for wind speed represent the 1:2 and 2:1 lines.

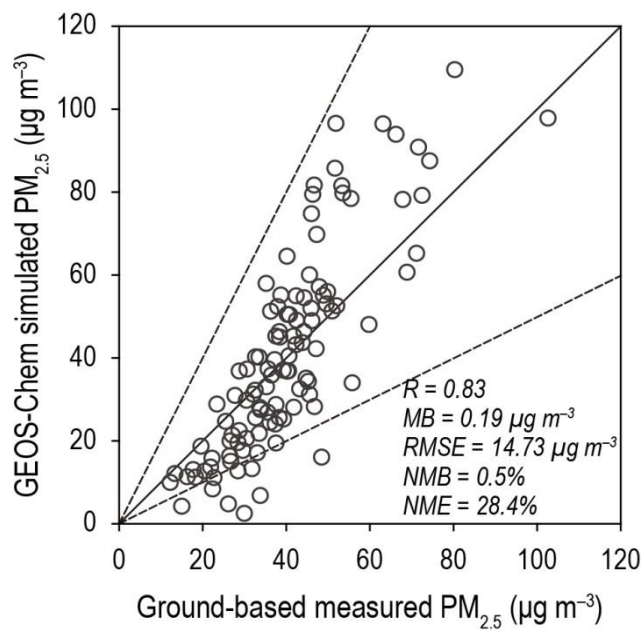

**Fig. S4.** The simulated surface PM$_{2.5}$ concentrations evaluation. The simulated surface PM$_{2.5}$ concentrations were compared to the ground-based measurements from MEE sites for GEOS-Chem grids over eastern China for the second half of 2016. Statistical parameters for all samples: correlation coefficient (R), mean bias (MB), root mean square error (RMSE), normalized mean bias (NMB), and normalized mean error (NME). The solid line represents the 1:1 line; dashed lines represent the 1:2 and 2:1 lines.

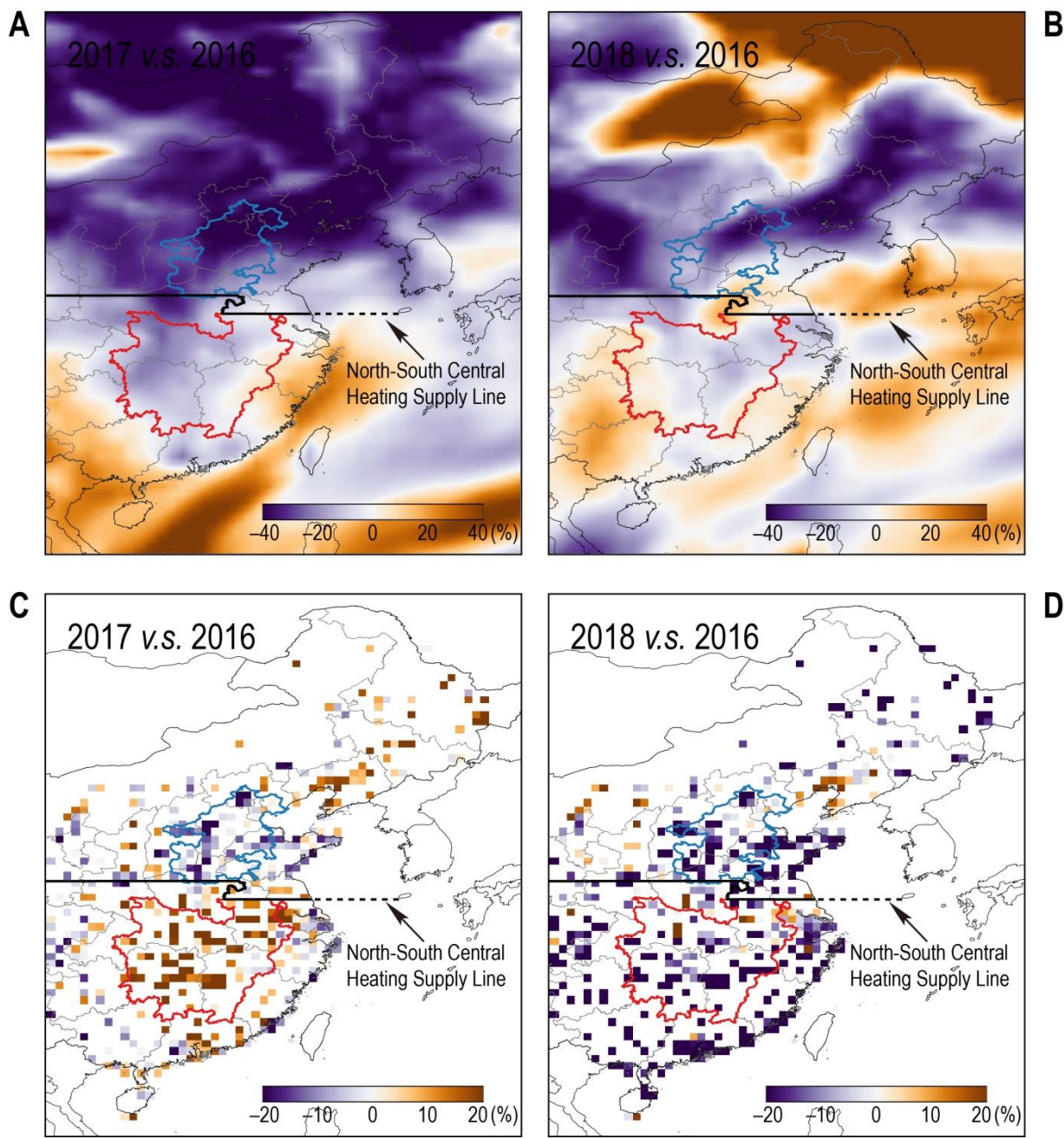

**Fig. S5.** Sensitivity examination for the meteorological effects with the nested GEOS-Chem. Maps show the relative changes of simulated PM$_{2.5}$ concentrations from the nested GEOS-Chem simulations (0.625 ° × 0.5 °, surface layer) for the heating period of 2017 (**A**) and 2018 (**B**), and of ground-based PM$_{2.5}$ concentrations with meteorological effects adjustments for the heating period of 2017 (**C**) and 2018 (**D**), on the basis of the same period of 2016. Symbols, see **Fig. 2** for an explanation.

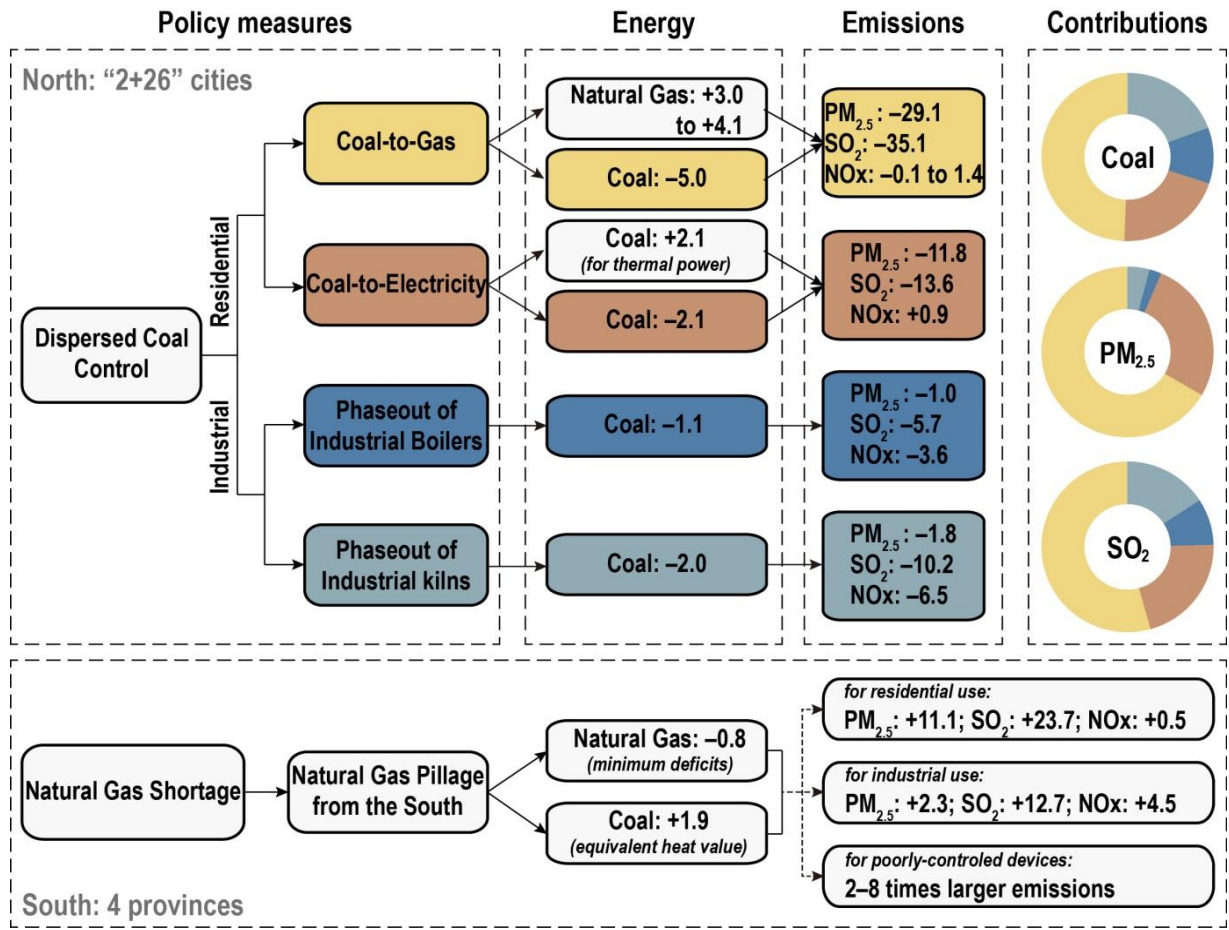

**Fig. S6.** The flowchart of emission changes for the studied regions in northern and southern China during the heating period 2016–2017. Colors denote the policy measures toward dispersed coal control implemented in the "2+26" cities in northern China. Negative values represent decreased energy consumption or air pollution emissions in the heating period 2017 and positive values are increases. Units of coal, million tonnes (Mt); natural gas, billion cubic metres in volume (bcm); and air pollution emissions, Gg.

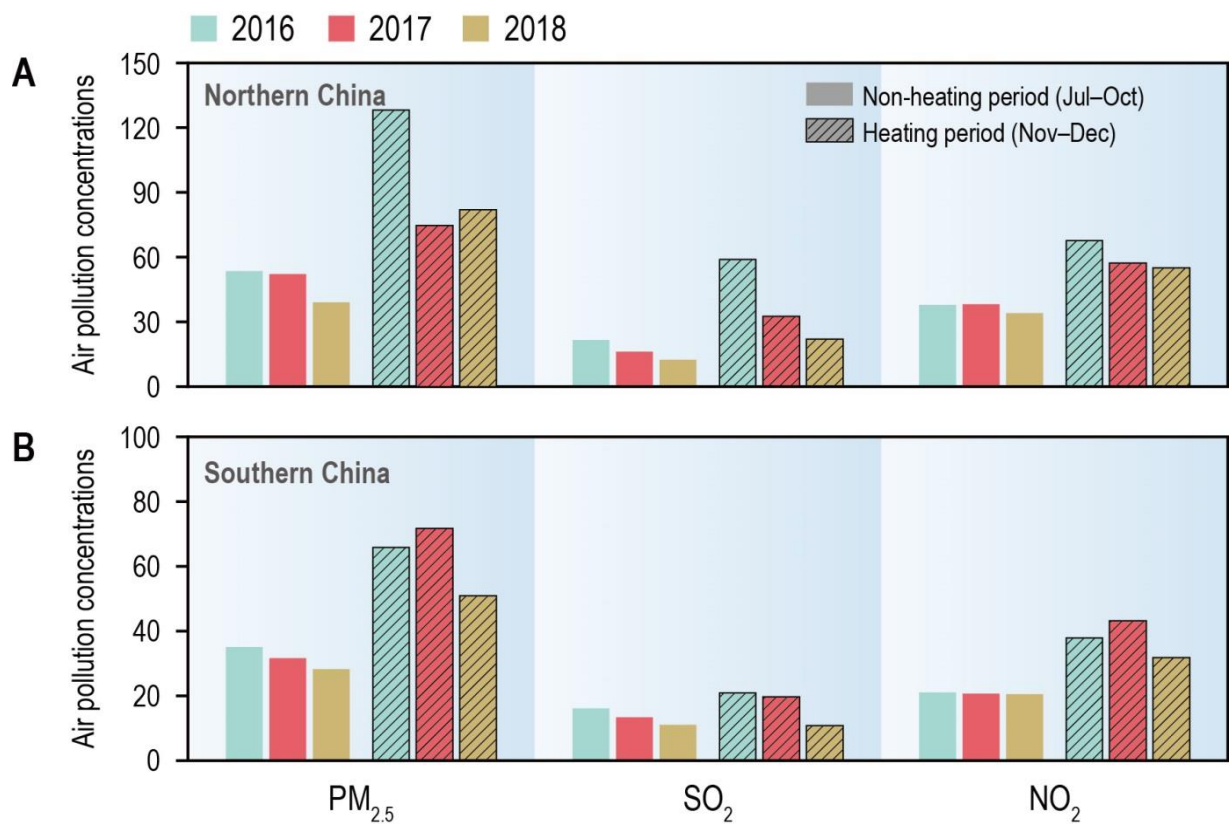

**Fig. S7.** Regional averaged air pollution concentrations during the non-heating and heating periods of 2016–2018. The three panels show the original measured average $PM_{2.5}$ (*left*), $SO_2$ (*middle*) and $NO_2$ (*right*) concentrations over the studied regions in northern (**A**) and southern (**B**) China, respectively.

Table S1. Emission factors for major air pollutants from coal and natural gas combustion in different facilities for the specific northern and southern regions

| Regions | Fuels | Facilities | Emission factors | | |
|---|---|---|---|---|---|
| | | | PM$_{2.5}$ | SO$_2$ | NOx |
| Northern | Dispersed coal | Residential stoves | 5.83 | 7.03 | 0.91 |
| | | Industrial boilers | 0.93 | 5.18 | 3.30 |
| | Bituminous coal | Power plant generators | 0.32 | 0.65 | 1.34 |
| Southern | Dispersed coal | Residential stoves | 5.83 | 12.45 | 0.91 |
| | | Industrial boilers | 1.21 | 6.71 | 3.30 |
| | Bituminous coal | Power plant generators | 0.40 | 1.25 | 1.74 |
| Both | Natural gas | Wall-mounted heaters | 0.00 | 0.00 | 1.46 |
| | | Industrial boilers | 0.00 | 0.00 | 2.08 |

Emission factors are sectoral averages adopted from the Multi-resolution Emission Inventory for China (MEIC) bottom-up inventory (http://www.meicmodel.org/) for the year of 2016, for the two specific regions of the "2+26" cities in northern China and the four inland provinces (Hubei, Hunan, Anhui, and Jiangxi) in southern China. Mass for NOx is calculated as NO$_2$. Units are g kg$^{-1}$ for coal and g m$^{-3}$ for natural gas.

Table S2. Emissions of air pollutants from MEIC inventory and the decreases estimated by MEIC during the heating period 2016–2017

| Regions | Species | Total emissions and decreases | | | Emission decreases by sector | | |
|---|---|---|---|---|---|---|---|
| | | 2016 | 2017 | Decreases | Industry | Power | Residential |
| Northern | $PM_{2.5}$ | 286.6 | 264.0 | 22.6 (7.9%) | 6.4 (5.0%) | 1.4 (7.0%) | 16.4 (13.6%) |
| | $SO_2$ | 469.5 | 355.4 | 114.1 (24.3%) | 66.5 (24.0%) | 29.8 (41.2%) | 18.2 (7.8%) |
| | NOx | 788.9 | 766.9 | 22.0 (2.8%) | –0.5 (–0.1%) | 22.4 (15.8%) | 1.9 (5.1%) |
| Southern | $PM_{2.5}$ | 246.9 | 232.9 | 14.0 (5.7%) | 6.2 (5.5%) | 0.5 (3.6%) | 8.4 (60.1 %) |
| | $SO_2$ | 338.6 | 263.9 | 74.7 (22.1%) | 50.6 (26.0%) | 10.7 (22.4%) | 13.8 (15.6%) |
| | NOx | 514.7 | 502.6 | 12.1 (2.4%) | 4.5 (2.1%) | 6.8 (9.2%) | 1.0 (4.8%) |

Emissions shown here are for the heating period (November–December) in 2016–2017. Transportation emissions account for less than 8% of the total changes during the studied period (not shown). Numbers in brackets provide the relative decreases in the heating period 2017 compared to 2016, and negative values mean increases. Units of Gg for emissions.